\documentclass[runningheads]{llncs}

\usepackage{soul}
\usepackage{microtype}
\usepackage{setspace}

\usepackage{amsmath}
\usepackage{amssymb}
\newcommand{\calH}{$\mathcal{H}$}

\usepackage{graphicx}
\usepackage{xcolor}
\usepackage{tabularx}
\usepackage{algorithm}
\usepackage{algpseudocode}
\AtBeginEnvironment{algorithmic}{\small}

\usepackage{subcaption}

\usepackage{float}

\usepackage{listings}
\PassOptionsToPackage{hyphens}{url}
\usepackage{hyperref}

\title{Finding Conservation Laws of Large Dynamical Systems with Tasks and Futures:\\A Case Study in Utilizing Dynamic Data Dependencies}
\titlerunning{A Case Study in Utilizing Dynamic Data Dependencies}\author{Rüdiger Nather}

\authorrunning{R. Nather}

\institute{
  University of Kassel\\
  Department of Electrical Engineering and Computer Science\\
  Kassel, Hesse, Germany\\
  \email{\{r.nather\}@uni-kassel.com}
}

\begin{document}

\maketitle

\begin{abstract}
  As parallel workloads grow in complexity, managing fine-grained data dependencies becomes a critical challenge.
  Futures offer a promising model for handling these dependencies, particularly in irregular algorithms, but they also come with the restriction of value-immutability.
  This immutability limits the ability to perform in-place memory updates, a necessity for high-performance linear algebra where memory recycling is paramount.

  In this paper, we address these limitations by introducing a new construct, \texttt{await\_delete}, which extends traditional future semantics to allow safe value reuse once consumers are finished.
  Building on this extension, we present a novel future-based algorithm for the block-wise inversion of dense, symmetric matrices, motivated by a recent algorithm for finding conservation laws of dynamical systems.

  We implement our approach in an extended version of Taskflow and evaluate it through strong-scaling experiments.
  Our results demonstrate that while futures incur significant overhead on smaller problem sizes, they achieve nearly linear scaling on large matrices.
  We analyze the amortization threshold and show that futures are a viable high-performance tool for large-scale linear algebra.

  \keywords{Dynamical Systems\and Conservation Laws \and Task-based Parallel Programming \and Data Dependencies \and Futures}
\end{abstract} 

\section{Introduction}\label{sec:introduction}
As the demand for computational power continues to grow, parallel programming models
must evolve to handle increasingly complex workloads.
Among these models, futures, awaitable placeholders for asynchronously computed values,
offer a compelling approach to managing data dependencies with fine granularity.
By allowing a program to only await the specific parts of a computation that are strictly necessary,
futures can effectively avoid induced dependencies that might otherwise stall execution.
This capability makes them an attractive candidate for parallelizing algorithms with complicated,
irregular dependency graphs;
especially those containing dynamic (data) dependencies which can only be observed at runtime.

However, the adoption of futures is not without trade-offs.
The runtime bookkeeping required to track asynchronous execution state incurs additional
memory allocations and management, which can become a bottleneck.
Furthermore, the standard semantics of futures typically require that values
remain constant once computed.
This immutability simplifies reasoning about concurrency but limits the flexibility of the model,
particularly in scenarios where in-place updates or memory recycling are essential for performance.

To understand where futures reside within the vast design space of dependency
management and parallelization, and to analyze how well they perform in practice,
it is necessary to not just evaluate them on synthetic benchmarks, but to also
examine a real-world problem characterized by a heavy computational load and complicated dependencies.

For this purpose, we focus on the parallelization of an algorithm for finding conservation laws of
dynamical systems, recently proposed by Mebrartie et al\cite{Mebratie2025mlcons}.
The most computationally expensive component of this algorithm involves the inversion of large, symmetric matrices.

To mitigate the substantial memory requirements of this approach, the original authors suggested the use
of hierarchical matrices (\calH-matrices).
\calH-matrices are recursively subdivided block matrices in which some leaf blocks are stored in a compressed form.
However, parallel algorithms for \calH-matrix arithmetic typically involve complicated dependency patterns
(see \cite{Saez2019exploiting,Nather2024futures}).
Interestingly, future-based algorithms have already been proposed specifically tailored for the
\calH-LU decomposition \cite{Nather2024futures,Nather2025futures},
demonstrating the utility of futures for this exact problem.

But, while \calH-matrices offer significant memory savings, we first need a baseline to determine how well
futures perform on dense matrices.
Otherwise, isolating the overhead of the future model would be difficult, as the 
computational gain from \calH-matrix arithmetic would skew the performance measurement.
Therefore, to isolate the overheads and benefits of the futures model itself,
we first restrict our investigation to the dense matrix case.
This provides a controlled baseline that captures the essential dependency structure
without the performance gain from \calH-techniques.

To apply futures effectively to this dense matrix inversion, we must address the limitation
of value-immutability,
specifically the inability to recycle memory buffers once their consumers have finished.
In high-performance linear algebra, the ability to update values in place is often crucial.
To bridge this gap, we introduce a new construct, \texttt{await\_delete}, which allows us to
reuse or change the value of a future once no other readers for this value exist.
This extension allows us to manage memory more directly, while retaining the dependency-handling
benefits of the futures model.

Building on this extension, we present a future-based algorithm for the block-wise inversion of matrices.
While we only focus on the fully dense case in this paper,
this algorithm serves as the foundation for future work.
By introducing case distinctions for the leaf-blocks, it can readily be extended to \calH-matrices.
We have implemented this algorithm in an extended version of
Taskflow\cite{Nather2025futures,Huang2021taskflow} and conducted a series of strong-scaling
experiments to determine the threshold at which the
overhead associated with futures amortizes relative to the problem size.

We summarize our contributions as follows:
\begin{itemize}
\item We extend the standard futures model to support value updates via the \texttt{await\_delete} construct and discuss the theoretical implications of this relaxation.
\item We propose a novel future-based algorithm for the block-wise inversion of symmetric matrices, designed to maximize parallelism through fine-grained dependency management.
\item We provide a implementation and analysis of the proposed algorithm. Our results demonstrate that futures can achieve linear scaling on large problem sizes, but suffer from a significant overhead in smaller problem sizes. We discuss potential causes and outline how to lower the amortization threshold.
\end{itemize}

The remainder of this paper is structured as follows: Section~\ref{sec:background} provides some background about the future model and the algorithm from Mebrartie et al.
Section~\ref{sec:algorithm} presents the block-wise inversion algorithm, and Section~\ref{sec:evaluation} discusses the implementation details and performance results.
Section~\ref{sec:relatedWork} summarizes related work and Section~\ref{sec:conclusion} presents our conclusions.

\section{Background}\label{sec:background}
This section provides the necessary background for this paper. 
We begin by defining the specific future model and semantics utilized in our implementation, drawing upon established work in \cite{Nather2024futures,Nather2025futures}. 
We then review how these futures are applied to manage the complex, recursive dependencies found in hierarchical matrix ($\mathcal{H}$-matrix) arithmetic. 
Finally, we outline the mathematical formulation for finding conservation laws in dynamical systems,
which serves as the primary driver for our case study.

\subsection{Futures for dynamic dependencies}\label{sec:background:futs}
We adopt the definition and implementation of futures described in~\cite{Nather2024futures} and~\cite{Nather2025futures}.
Futures act as placeholders for values, featuring distinct writer and reader sides.
Specifically, each placeholder is associated with a single \textit{promise} (used to write the value)
and one or more \textit{futures} (used to read the value).
The promise is fulfilled exactly once; upon fulfillment, all associated futures become ready immediately.

Promises and futures are identified by their name and type, and can be stored in standard data structures such as arrays.
The element type is unrestricted, enabling advanced constructs such as futures of futures.
When passed as parameters, futures are duplicated as needed, while promises remain unique and are always moved (and never copied).
Moving a promise transfers the responsibility for filling it to the recipient task.
A task is ready to execute only when all futures it depends on have been fulfilled.

Regarding memory management, our implementation utilizes a reference-counting scheme.
When all futures referencing a specific placeholder go out of scope,
the placeholder and its underlying memory are automatically deallocated, providing a limited form of automatic memory reclamation.

We will use the general notion that \texttt{spawn}ing a task will return a future for the result of the task. A task can also be spawned \texttt{into} a promise, thereby binding the result of that task to a pre-existing promise.
The code in Listing~\ref{lst:fib} demonstrates how recursive computations and dependencies are expressed using the promise/future model.
\begin{itemize}
\item[\texttt{add}:] When used to create a task, takes two integers, computing their sum,
  and fulfills the output promise \texttt{p} via \texttt{p.set(i+j)}.

\item[\texttt{fib}:] Computes the $n$-th Fibonacci number $F(n)$ and fulfills promise \texttt{p}:
  \begin{itemize}
  \item Base case: For $n < 2$, fulfills \texttt{p} immediately with $1$.

  \item Recursive case:\begin{enumerate}
    \item Spawn \texttt{fib(n-1)}, obtaining future \texttt{f1} for $F(n - 1)$.
    \item Create promise \texttt{p2} for $F(n-2)$ and obtain its future \texttt{f2}.
    \item Spawn \texttt{add} with \texttt{f1}, \texttt{f2}, and output promise \texttt{p} (scheduled only after both futures are ready)
    \item Compute $F(n-2)$ directly via \texttt{fib(n-2, p2)} (synchronous in this example).
    \end{enumerate}
  \end{itemize}
\end{itemize}

\begin{minipage}{\textwidth}
\begin{lstlisting}[caption={ Example for usage of programming model},label=lst:fib]
int add(int i, int j) { return i+j; }
void fib(int n, promise<int> p) {
  if (n < 2) {
    p.set(1);
  } else {
    future<int> f1 =
    spawn fib(n-1, make_promise());
    promise<int> p2 = make_promise();
    future<int> f2 = p2.get_future();
    spawn into(p) add(f1, f2);
    fib(n-2, p2);
  }
}
\end{lstlisting}
\end{minipage}

\subsection{Futures for recursive block-matrices}\label{sec:background:hmat}
The application of futures to $\mathcal{H}$-matrix arithmetic, as detailed in~\cite{Nather2024futures}, leverages the recursive nature of the data structure.
Non-elementary tasks recursively spawn other non-elementary tasks as they traverse the matrix hierarchy, eventually reaching the leaf level where elementary tasks are executed. Elementary tasks perform the actual, numerical work; in our implementation, this will be done by an external library.
Futures are used to track dependencies throughout this process.

As tasks execute, several pieces of information must be communicated between them to manage dependencies and data flow:
\begin{itemize}
\item[(i1)] Has the task been completed?
  For non-elementary tasks, this means that all child tasks have been spawned, whereas for elementary tasks it indicates that the associated matrix block has been updated.
  The cases are combined, since a (non-elementary) task generally does not know whether another task is elementary or not.

\item[(i2)] Is the update of the associated matrix block complete?
  (This is only relevant for non-elementary tasks, where (i2) differs from (i1))

\item[(i3)]  Contents of the (updated) matrix block
  (This is only relevant for elementary tasks)

\item[(i4)]  Futures of all child tasks (This is only relevant for non-elementary tasks)
\end{itemize}

To facilitate this communication, we define the recursive node structure in Listing~\ref{lst:type}.
The \texttt{rec\_matrix\_node} struct contains a \texttt{data} future for storing the memory address of leaf blocks and a boolean flag \texttt{leaf} to distinguish between internal and leaf nodes.
The \texttt{matrix\_t} type represents the underlying matrix type used by the external library mentioned above.
The array of futures \texttt{subnodes} contains the futures for child tasks in non-leaf nodes.
Finally, the \texttt{done} future signals the completion of the operation associated with the node.

It is important to distinguish the roles of these futures: \texttt{data} becoming ready signifies that the memory address for the block is known (essential for non-elementary tasks to schedule children), whereas \texttt{done} becoming ready signifies that the computation on that block (and all subnodes) has finished.
Which task is responsible for setting them depends on the algorithm being implemented.

\begin{minipage}{\textwidth}
  \begin{lstlisting}[caption={Type for the Recursive Subdivision},label=lst:type]
struct rec_matrix_node {
   future<matrix_t*> data;
   bool leaf;
   array<future<rec_matrix_node>> subnodes;
   future<void> done;
}
\end{lstlisting}
\end{minipage}

\subsection{Finding conservation laws by Kernel Ridge Regression}\label{sec:background:conslaw}
The primary motivation for our matrix inversion algorithm stems from the problem of identifying conservation laws in dynamical systems.

Consider a dynamical system defined by $\dot{\mathbf{x}} = \mathbf{f}(\mathbf{x})$, where $\mathbf{x} \in \mathbb{R}^n$.
A conservation law is a scalar function $C(\mathbf{x})$ that remains constant along the trajectories of the system, satisfying $\nabla C(\mathbf{x}) \cdot \mathbf{f}(\mathbf{x}) = 0$.

As proposed by Mebrartie et al.~\cite{Mebratie2025mlcons}, this problem can be formulated as a regression task using Kernel Ridge Regression (KRR).
Given a set of trajectory data $\mathcal{D} = \{(\mathbf{x}_{(i,j)})_{i=1}^N\}_{j=1}^M$, where $N$ is the number of samples per trajectory and $M$ is the number of trajectories, the objective is to find a function $C(\mathbf{z})$ that minimizes the deviation from constancy along each trajectory.

The conservation law $C(\mathbf{z}) = \sum_{i=1}^{NM} \alpha_i \varphi(\mathbf{z}, \mathbf{x}_i)$ is then expressed as a linear combination of kernel functions \mbox{$\varphi(\mathbf{x}, \mathbf{y}) = (\langle \mathbf{x}, \mathbf{y} \rangle - c)^d$}, 
where $\boldsymbol{\alpha} \in \mathbb{R}^{NM}$ are the coefficients to be determined.

To find the coefficients $\boldsymbol{\alpha}$, one must solve a quadratic optimization problem.
This involves the inversion of the regularized kernel matrix $\mathbf{K} - \lambda \mathbf{I} \in \mathbb{R}^{NM \times NM}$, 
where $\lambda > 0$ is a regularization parameter and the symmetric matrix $\mathbf{K}$ is computed from the data points $\mathcal{D}$.

The authors note that execution time is dominated by the repeated inversion of kernel matrices.
Since matrix size depends on the number of training samples ($N\cdot M$), these matrices can become very large.
Furthermore, identifying optimal parameters via cross-validation requires solving this inverse problem multiple times for different matrix configurations.

\section{Algorithm}\label{sec:algorithm}
Standard futures enforce value immutability, preventing in-place updates and forcing new memory allocations for every computation step.
To address this, we introduce \texttt{await\_delete}, a primitive that allows values to be reused or overwritten once it is
guaranteed that no other tasks will read them.

\subsection{The \texttt{await\_delete}-Construct}
The \texttt{await\_delete} construct takes an existing future $f$ as input and returns a new future $f'$.
The promise associated with $f'$ is implicit and tied to the lifetime of the original future $f$.
Specifically, $f'$ is fulfilled only when the reference count of $f$ drops to zero, indicating that all consumers of the original value have released their handles.

From the perspective of a task waiting on $f'$, the behavior is identical to waiting on a standard future.
The fulfillment of $f'$ then signals not just that data is available, but that the \textit{exclusive ownership} of the underlying memory has been reclaimed.
This allows the consumer of $f'$ to safely treat the memory as mutable and reuse it for subsequent operations without allocating new buffers.

It is important to note that \texttt{await\_delete} can only be invoked once per future, as it assumes that the reference count is sufficient to track exclusivity.
Provided this invariant holds, scheduling tasks according to their future-based dependencies (as outlined in Subsection~\ref{sec:background:futs}) guarantees deterministic execution order.
However, as with standard parallel floating-point arithmetic, bit-wise reproducibility is not guaranteed due to the non-associative nature of floating-point operations.

Furthermore, the \texttt{await\_delete} construct composes safely within arbitrary dependency graphs, including nested structures.
Whether $f$ is the result of a simple computation or a complex chain of nested futures, $f'$ is fulfilled only when the specific future $f$ becomes unreachable.
This local property ensures that \texttt{await\_delete} can be applied to any node in a graph without introducing race conditions or deadlocks,
provided the write-once constraint of the promise is maintained and \texttt{await\_delete} is invoked at most once per future.

\subsubsection{Application to Leaf Blocks:}
In the context of dense matrix arithmetic, elementary tasks operate on individual matrix blocks.
Without \texttt{await\_delete}, a sequence of operations (e.g., scaling followed by addition) would typically require allocating a temporary buffer for the intermediate result.
With \texttt{await\_delete}, we can chain these operations on the same memory region.

Consider a block $B$ represented by a future $f_B$.
A task $T_1$ reads $f_B$ and produces a new block $B'$, represented by future $f_{B'}$.
If a subsequent task $T_2$ requires $B'$ as input, it typically waits on $f_{B'}$.
However, if $T_2$ is intended to update $B'$ in-place, we instead pass $f'_{B'} = \texttt{await\_delete}(f_{B'})$ to $T_2$.
This ensures that $T_2$ can only be executed once $T_1$ is finished and no more consumers for the result of $T_1$ exist, thereby allowing us to reuse the memory where block $B'$ is stored.
Note that this extends to (de)allocation as well: allocating space for the block is fulfilling the corresponding promise, whereas deallocation simply becomes the last task in the sequence of operations.

\subsubsection{Application to Recursive Structures:}\label{sec:recStruct}
In recursive data structures like the one shown in Listing~\ref{lst:type}, operations often involve traversing the tree and modifying nodes.
A critical challenge arises when multiple sequential operations must be performed on the same structure (e.g., a forward factorization pass followed by a backward substitution pass). In this case it needs to be ensured that the elementary tasks on the leafs are sequenced correctly.

Consider a recursive node and two sequential operations: $T_1$ and $T_2$.
If these non-elementary tasks are scheduled using standard futures for the child nodes, a race condition may occur.
Specifically, the parent task responsible for $T_2$ might finish spawning its children before the parent task for $T_1$ has finished spawning \textit{its} children.
This would allow elementary tasks from $T_2$ to execute before elementary tasks from $T_1$ on the same data,
inverting the intended sequential dependencies.

Passing an \texttt{await\_delete}d future for a child node solves this issue by enforcing strict sequentialization at the spawner level.
When a parent task completes, that is, finishes spawning tasks, it drops its handles to the child futures.
By having the next task wait for \texttt{await\_delete(child\_future)}, we ensure that the next operation cannot begin spawning until the previous operation has finished spawning.
This effectively ``peels back'' the recursive layers one by one, ensuring that task spawns cannot overtake each other.
Crucially, this does not require waiting for the entire sub-tree of tasks to complete; we only synchronize the boundary between the sequential passes at each layer individually.

\subsection{Blockwise Inversion of Symmetric Matrices}\label{sec:invert}
We apply the \texttt{await\_delete} construct to the problem of inverting a symmetric matrix using a standard block-wise recursive approach.
Given a symmetric matrix $A$ partitioned into four blocks, the inverse $A^{-1}$ can be computed using the following block decomposition:

\begin{equation*}
A^{-1} = 
\begin{pmatrix}
A_{11} & A_{12} \\
A_{21} & A_{22}
\end{pmatrix}^{-1}
=
\begin{pmatrix}
S_{11} & S_{12} \\
S_{21} & S_{22}
\end{pmatrix},
\end{equation*}
where the intermediate steps are defined as:
\vspace{-6pt}\[T_{11} = \text{inv}(A_{11}),\;\;\;  T_{12} = A_{21} T_{11},\;\;\; T_{22} = A_{22} - T_{12} A_{21} \]
\[S_{22} = \text{inv}(T_{22}),\;\;\;  S_{12} = S_{21}^T,\;\;\;  S_{21} = - T_{22} T_{12},\;\;\;  S_{11} = T_{11} - S_{21}^T S_{12}\]

Note that each operation overwrites the respective block; the notation above distinguishes the intermediate values ($T$, $S$) to clarify the data flow.

Translating this into the future-model is then straightforward; each overwrite happens on an \texttt{await\_delete}d future, where the result is passed to the subsequent operations.
Algorithm~\ref{alg:invert} details this process.

\begin{algorithm}[ht]
\caption{Blockwise Matrix Inversion}\label{alg:invert}
\begin{algorithmic}[1]
\Function{Invert}{$node$}
    \If{$node$ is leaf}
        \State $block \gets \texttt{await\_delete}(node.data)$ \Comment{Reuse memory of $node.data$}
        \State $done \gets \textsc{InvertBlock}(block)$
        \State \Return $\textsc{RecNode}(block, done)$
    \Else
    \State $T_{11} \gets \textsc{Invert}(\texttt{await\_delete}(node.sub[0]))$ \Comment{Reuses $A_{11}$ memory}
        \State $T_{12} \gets \textsc{MatMul}(node.sub[2], T_{11}, \texttt{await\_delete}(node.sub[1]))$\phantom{mmmmmn}
        \hspace*{5cm}\Comment{Reuses $A_{12}$ memory}
        \State $T_{22} \gets \textsc{MatMul}(-1.0, T_{12}, node.sub[2], 1.0, \texttt{await\_delete}(node.sub[3]))$
        \State $S_{22} \gets \textsc{Invert}(\texttt{await\_delete}(T_{22}))$ \Comment{Reuses $A_{22}$ memory}
        \State $S_{21} \gets \textsc{MatMul}(-1.0, S_{22}^{-1}, T_{12}, 0.0, \texttt{await\_delete}(node.sub[2]))$\phantom{mmmmm}
        \hspace*{5cm}\Comment{Reuses $A_{21}$ memory}
        \State $S_{12} \gets \textsc{Transpose}(S_{21}, \texttt{await\_delete}(T_{12}))$\Comment{Reuses $A_{12}$ memory}
        \State $S_{11} \gets \textsc{MatMul}(-1.0, S_{12}^T, S_{21}, 1.0, \texttt{await\_delete}(T_{11}))$\phantom{mmmmmmmmmn}
        \hspace*{5cm}\Comment{Reuses $A_{11}$ memory}
        \State $done \gets \textsc{Wait}(S_{22}.done, S_{21}.done, S_{12}.done, S_{11}.done)$
        \State \Return $\textsc{RecNode}(S_{11}, S_{12}, S_{21}, S_{22}, done)$
    \EndIf
\EndFunction
\end{algorithmic}
\end{algorithm}

The function \textsc{Invert} takes a recursive node representing a matrix block.
If the node is a leaf, we simply invoke an external routine to invert the dense block in-place.

For non-leaf nodes, the algorithm follows the dependencies defined in the block decomposition.
We spawn tasks to compute the intermediate components ($T_{11}, T_{12}, T_{22}$) and tasks for the final inverse blocks ($S_{11}, \dots, S_{22}$) sequentially.

Crucially, each step reuses the memory of a specific input block:
\begin{itemize}
\item \textbf{Line 5:} We recursively invert the top-left block $A_{11}$. The result $T_{11}$ is stored directly in the memory previously occupied by $A_{11}$.
\item \textbf{Line 6:} We compute $T_{12} = A_{21} T_{11}$. The result will overwrite the top-right block $A_{12}$.
\item \textbf{Line 7:} We compute the Schur complement $T_{22} = A_{22} - T_{12} A_{21}$. This is a fused multiply-add operation that will overwrite the bottom-right block $A_{22}$.
\item \textbf{Line 8:} We recursively invert the Schur complement $T_{22}$ to obtain $S_{22}$, reusing the same memory slot.
\item \textbf{Line 9:} We compute $S_{21} = -S_{22} T_{12}$. This will overwrite the bottom-left block $A_{21}$.
\item \textbf{Line 10:} We compute $S_{12}$ as the transpose of $S_{21}$. This reuses the memory currently holding $T_{12}$ (which was previously $A_{12}$).
\item \textbf{Line 11:} Finally, we compute $S_{11} = T_{11} - S_{21}^T S_{12}$. This will overwrite the memory holding $T_{11}$ (which was originally $A_{11}$).
\end{itemize}

While standard block-wise algorithms overwrite input blocks to save memory, traditional futures would necessitate allocating new memory for every intermediate result, incurring significant overhead. The \texttt{await\_delete} construct resolves this by enabling precise memory reuse sequencing. It signals when a block is no longer needed by dependent tasks, allowing its memory to be safely reclaimed for subsequent operations. This creates a pipeline that recycles memory exactly between computational phases, maintaining parallelism while minimizing footprint as described in Subsection~\ref{sec:recStruct}.

Note that this algorithm does not make any assumptions about the storage layout of the leaf blocks. Indeed, the dependency management with futures allows the computation to flow downward, ultimately overwriting each (leaf) block without the tasks higher up in the recursion hierarchy needing to know \emph{where} they are actually stored.
By the same token, it also generalizes to both operations on non-square matrices, as well as the inversion of non-symmetric matrices.
However, while inverting non-symmetric matrices will require additional working space, utilizing the \texttt{await\_delete} construct allows us to bound this overhead.
Specifically, we can reuse the memory of the input matrix to store the result (as above), but also of the additional working space,
thereby limiting the total auxiliary memory requirement to at most half the size of the matrix.

\section{Evaluation}\label{sec:evaluation}
To analyze the performance of our future-based matrix inversion, we conducted a series of strong-scaling experiments.
We varied both the total size of the matrix and the granularity of the tasks by adjusting the size of the leaf blocks.
Specifically, we tested three matrix sizes: $2^{14} \times 2^{14}$, $2^{15} \times 2^{15}$, and $2^{16} \times 2^{16}$.
For each matrix size, we varied the leaf block size such that the ratio between the matrix size and the block size was $512$, $256$, and $128$.
It is important to note that the total numerical work for each matrix size remains constant across different block sizes; the only variable is the granularity of the parallel tasks.

\subsection{Runtime Analysis}
Figure~\ref{fig:runtime} presents the execution time measurements.
Each subfigure corresponds to one matrix dimension, with the three curves representing the different block size ratios.
The results confirm that parallelism is effectively exploited, as execution time decreases with increasing core count for all configurations.
However, the impact of task granularity is significant.
In Subfigure~\ref{fig:runtime:0} ($2^{14} \times 2^{14}$), we observe that finer granularity (smaller block sizes, corresponding to higher ratios) worsens performance.
Specifically, the configuration with a ratio of $512$ (smallest blocks) exhibits notably higher runtimes compared to the coarser configurations.
For the larger matrix sizes ($2^{15}$ and $2^{16}$), the performance gap between block sizes $64$ and $128$ (ratios $256$ and $128$) is less pronounced, though the trend of coarser granularity yielding better performance persists.

\subsection{Speedup Analysis}
To better quantify the efficiency of the parallelization, Figure~\ref{fig:speedup} depicts the computed speedup curves.
Each subfigure corresponds to a ratio between matrix size and block size.
For the largest matrix size ($2^{16} \times 2^{16}$), the speedup curves are nearly identical across all block size ratios (Subfigure~\ref{fig:speedup:2}).
This suggests that for sufficiently large problems, the overhead associated with fine-grained task management becomes negligible compared to the computational work.
For the intermediate matrix size ($2^{15} \times 2^{15}$), the optimal performance is achieved at a ratio of $256$ (Subfigure~\ref{fig:speedup:1}).
Here, the finer granularity of ratio $512$ leads to a noticeable degradation in speedup, while the coarser ratio of $128$ performs similarly to the optimal case.
Interestingly, for the smallest matrix size ($2^{14} \times 2^{14}$), the speedup at ratio $128$ converges with the performance of the $2^{15}$ matrix at the same ratio, indicating that the task granularity is sufficiently coarse to amortize the overhead even for smaller problem sizes.

\subsection{Discussion of Overheads}The performance degradation at fine granularities points to a runtime bottleneck rather than an algorithmic flaw.
We attribute this to the current future implementation, which uses a locking scheme (atomic compare-and-swap) for reference counting.
With many small tasks, congestion arises as numerous threads simultaneously attempt to read or fulfill futures, causing significant synchronization overhead.
This effect is most pronounced when the task creation-to-computation time ratio is high.
As matrix size increases, the computation-to-communication ratio improves, diminishing the relative cost of lock contention and closing the performance gap for the $2^{16}$ case.

\begin{figure}[H]
  \centering
  \begin{subfigure}{\textwidth}
    \caption{Runtime breakdown for matrix size $2^{14}$. \label{fig:overhead:s14}}
    \includegraphics[width=\textwidth]{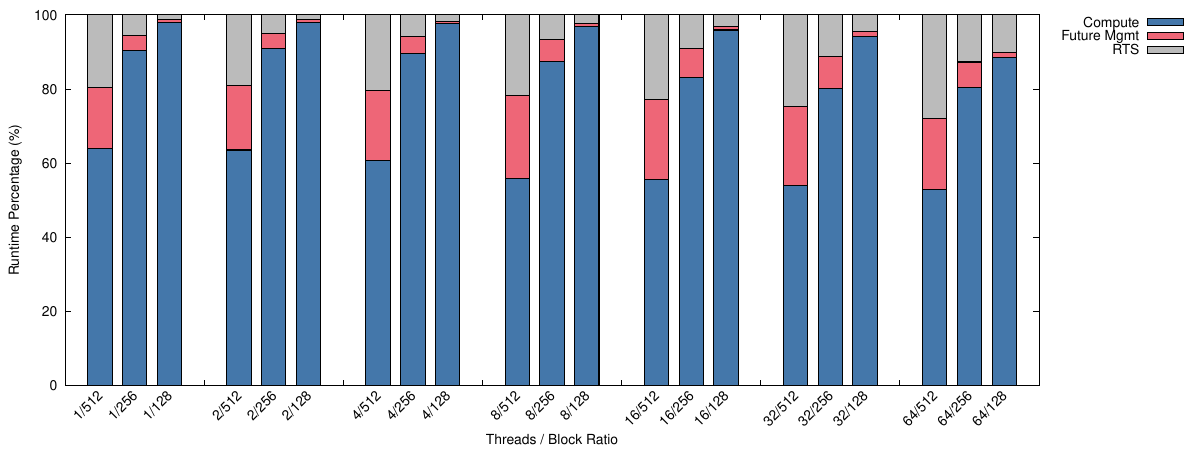}
  \end{subfigure}
  
  \begin{subfigure}{\textwidth}
    \caption{Runtime breakdown for matrix size $2^{15}$. \label{fig:overhead:s15}}
    \includegraphics[width=\textwidth]{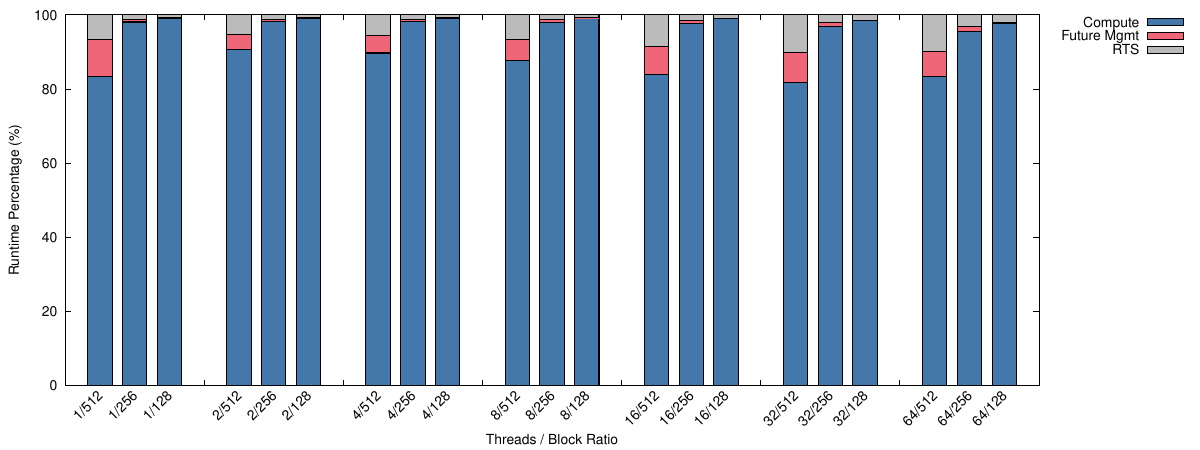}
  \end{subfigure}
  
    \caption{Bars are grouped by thread count, with each cluster showing the impact of different block size ratios. The blue segment represents compute time, the red segment represents time spent in future management.}
    \label{fig:overhead_breakdown}
\end{figure}

To validate our hypothesis that lock contention is the primary bottleneck, we profiled the execution to separate the time spent in actual computation from the time spent managing futures.
Figure~\ref{fig:overhead_breakdown} provides a detailed breakdown of the runtime for matrix sizes $2^{14}$ (Subfigure~\ref{fig:overhead:s14}) and $2^{15}$ (Subfigure~\ref{fig:overhead:s15}), categorizing the time into three components: \textit{compute} time (actual numerical computation), \textit{future management} (every operation that touches the shared state of futures), and \textit{RTS} overhead.

The data clearly illustrates the trade-off between granularity and overhead.
For configurations with fine granularity (high ratio $r$), the time spent on future management (red) constitutes a significant portion of the total runtime.
This confirms that the cost of managing dependencies dominates when tasks are small.
Conversely, as the block size increases (lower ratio $r$), the proportion of time spent on future management drops and the application becomes compute-bound.

Comparing this approach to a traditional threading model like OpenMP would likely show lower overheads for the dependency management.
However, such models lack the dynamic flexibility required for our target application, specifically the handling of irregular, runtime-generated dependencies inherent to $\mathcal{H}$-matrix arithmetic and the conservation law problem (Subsections~\ref{sec:background:hmat} and~\ref{sec:background:conslaw}).
Thus, the observed overhead represents the cost of this increased expressiveness.

\noindent
\hspace{-0.09\textwidth}
\begin{minipage}{0.58\textwidth}
\begin{figure}[H]
  \centering
  \begin{subfigure}{\textwidth}
    \caption{Matrix size: $16384\times 16384$\label{fig:runtime:0}}
    \includegraphics[width=\textwidth]{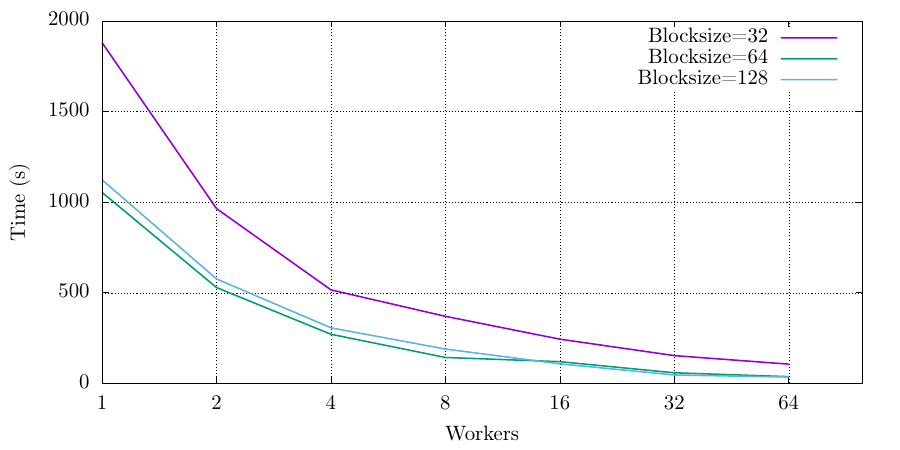}
  \end{subfigure}
  
  \begin{subfigure}{\textwidth}
    \caption{Matrix size: $32768\times 32768$\label{fig:runtime:1}}
    \includegraphics[width=\textwidth]{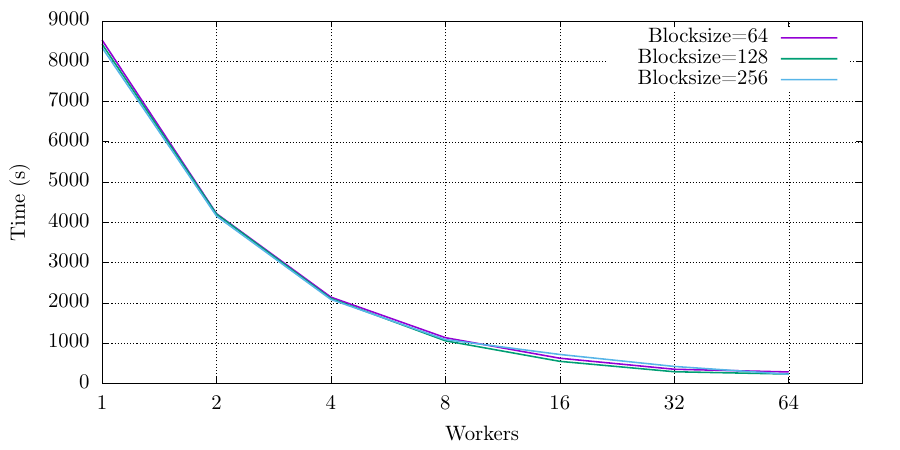}
  \end{subfigure}
  
  \begin{subfigure}{\textwidth}
    \caption{Matrix size: $65536\times 65536$\label{fig:runtime:2}}
    \includegraphics[width=\textwidth]{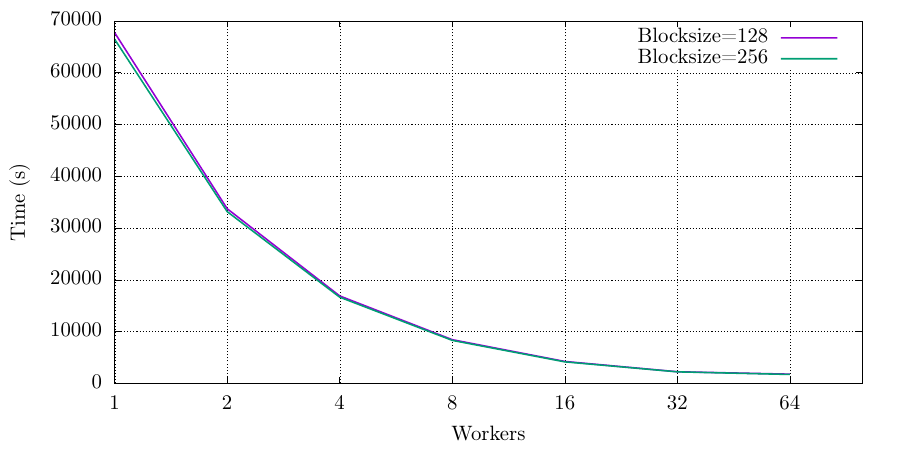}
  \end{subfigure}
  \caption{Running Times\label{fig:runtime}}
\end{figure}
\end{minipage}
\begin{minipage}{0.58\textwidth}
\begin{figure}[H]
  \centering
  \begin{subfigure}{\textwidth}
    \caption{Size-Blocksize ratio: $512$\label{fig:speedup:0}}
    \includegraphics[width=\textwidth]{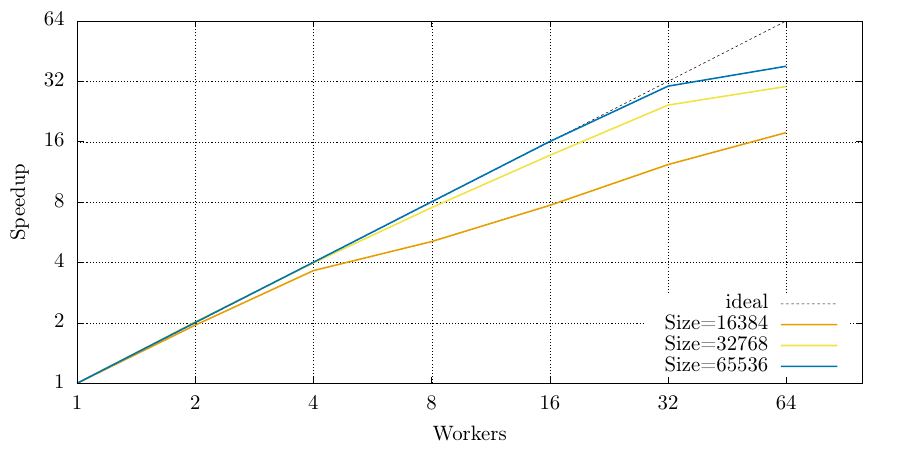}
  \end{subfigure}
  
  \begin{subfigure}{\textwidth}
    \caption{Size-Blocksize ratio: $256$\label{fig:speedup:1}}
    \includegraphics[width=\textwidth]{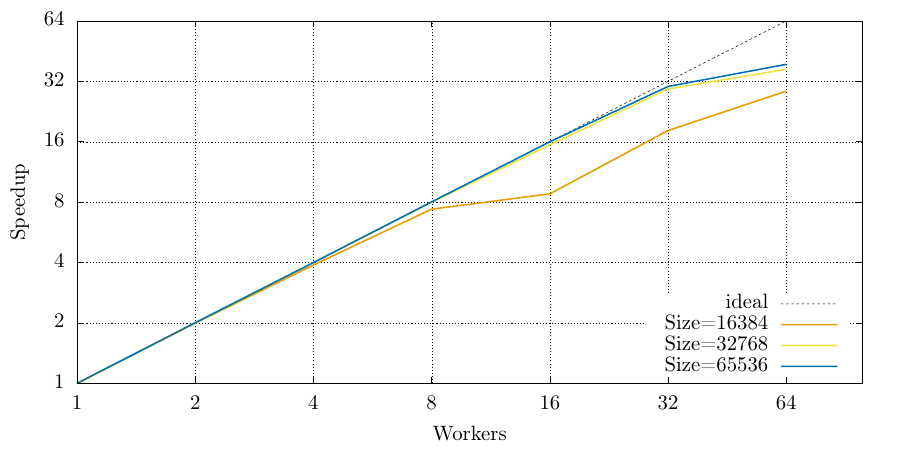}
  \end{subfigure}
  
  \begin{subfigure}{\textwidth}
    \caption{Size-Blocksize ratio: $128$\label{fig:speedup:2}}
    \includegraphics[width=\textwidth]{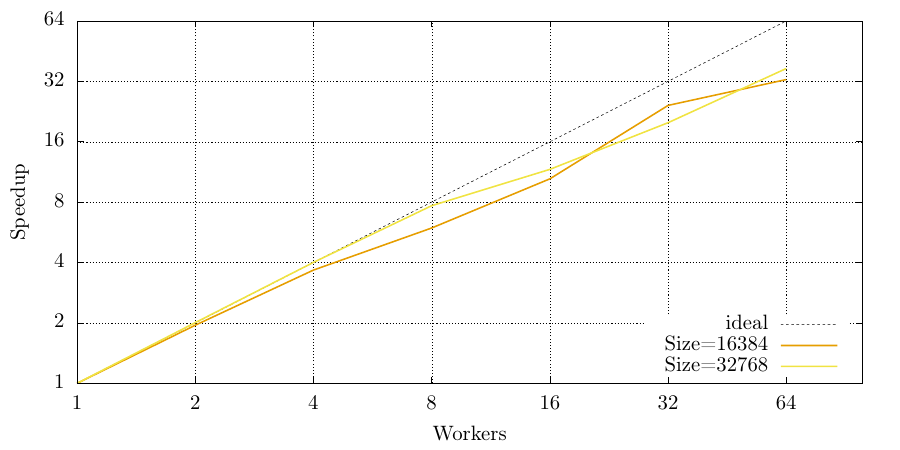}
  \end{subfigure}
  \caption{Speedup\label{fig:speedup}}
\end{figure}
\end{minipage}

\section{Related Work}\label{sec:relatedWork}

Taskflow~\cite{Huang2021taskflow} already provides a notion of composability of dependencies through module tasks,
but these are static programming constructs designed as a design-time abstraction and lack the flexibility required for the dynamic and recursive task generation inherent in our matrix inversion algorithm.

The data-versioning approach in Superglue~\cite{Tillenius2015superglue} is similar to our approach.
However, it requires manual management of both data partitioning and versioning,
whereas our future-based approach follows the recursive nature of subdivided matrices more closely.
Together with our new \texttt{await\_delete} construct, this gives an implicit data versioning which avoids the
additional development effort.

While Parsec~\cite{Hoque2017parsec} automatically orders tasks to prevent premature data overwrites,
its dependency model does not easily accommodate dependencies between non-sibling tasks, which are common in \calH-arithmetic.

\section{Conclusion}\label{sec:conclusion}
This work demonstrates that futures offer a compelling solution for managing complex, irregular dependency patterns in parallel algorithms.
Our empirical analysis reveals a critical trade-off: while futures incur significant overhead for small problem sizes, they achieve nearly linear scaling on large matrices.

The \texttt{await\_delete} construct is essential to this performance profile.
By enabling safe memory reuse at the task level, it allows to maintain the memory efficiency of in-place linear algebra while preserving the dependency management benefits of futures.
This facilitates seamless integration with external numerical libraries, as memory management is handled entirely within the task model.
While \texttt{await\_delete} provides a pragmatic solution, we view it as a temporary extension; future work should aim to integrate memory ownership semantics more natively into the future model to eliminate the need for ad-hoc constructs.

Looking ahead, we plan to extend our approach to hierarchical matrices to fully implement the pipeline for finding conservation laws in dynamical systems.
Additionally, we aim to optimize the runtime to reduce the overhead of the current implementation and lower the amortization threshold.

\bibliographystyle{splncs04}

\end{document}